\documentclass[11pt]{article}
\usepackage{amsmath,amssymb,cite}
\usepackage{graphicx}
\makeatletter
\@addtoreset{equation}{section}
\makeatother

\newcommand{\ba}{\begin{eqnarray}}
\newcommand{\ea}{\end{eqnarray}}
\def\half{\frac{1}{2}}
\def\ben{\begin{equation}}
\def\een{\end{equation}}
\def\bea{\begin{eqnarray}}
\def\eea{\end{eqnarray}}
\def\be{\begin{equation}}
\def\ee{\end{equation}}

\def\nn{\nonumber}
\def\p{\partial}

\def \ket {| \Psi \rangle}

\def \uket {| \uparrow \rangle}
\def \dket {| \downarrow \rangle}
\def \uuket {| \uparrow  \uparrow \rangle}
\def \udket {| \uparrow  \downarrow \rangle}
\def \duket {| \downarrow \uparrow\rangle}
\def \ddket {| \downarrow \downarrow\rangle}
\def \cH {{\cal H}}

\def\crampest{\medmuskip = 1mu plus 1mu minus 1mu}
\def\uncramp{\medmuskip = 4mu plus 2mu minus 4mu}

\def\ben{\begin{equation}}
\def\een{\end{equation}}
\def\bea{\begin{eqnarray}}
\def\eea{\end{eqnarray}}
\def \nn {\nonumber}

\def \bx {{\bf x}}
\def \p {\partial}

\def\half {\frac{1}{2}} 
\def\ft#1#2{{\textstyle{\frac{\scriptstyle #1}{\scriptstyle #2} } }}
\def\fft#1#2{{\frac{#1}{#2}}}

\def\R{\mathbb{R}}
\def\C {\mathbb{C}}
\def\Z{{\mathbb{Z}}}
\def\CP{\mathbb{C}\mathbb{P}}
\def\RP{\mathbb{R}\mathbb{P}}
\def\im{{\rm i\,}}
\def\ep{{\epsilon}}

\def\del{{\partial}}

\def\sech{{{\rm sech\,}}}

\newcommand{\hoch}[1]{$\, ^{#1}$}

\newcommand{\auth}{\Large\bf{M. Cveti\v c\hoch{1,6}, 
G.W. Gibbons\hoch{2,1,3,4}
and C.N. Pope\hoch{5,2}}}

\begin{document}

\begin{flushright}
\hfill {
UPR-1273-T\ \ \ 
MI-TH-1526 \ \ \
}\\
\end{flushright}

\begin{center}

{\LARGE{\bf Compactifications of Deformed Conifolds, Branes and the 
Geometry of Qubits}}

\vspace{10pt}
\auth

\large

\vspace{7pt}{\hoch{1}\it Department of Physics and Astronomy,\\
University of Pennsylvania, Philadelphia, PA 19104, USA}

\vspace{3pt}{\hoch{2}\it DAMTP, Centre for Mathematical Sciences,\\
 Cambridge University, Wilberforce Road, Cambridge CB3 OWA, UK}

\vspace{3pt}{\hoch{3}\it  
Laboratoire de Math\'ematiques et Physique Th\'eorique  CNRS-UMR
7350 \\
F\'ed\'eration Denis Poisson, Universit\'e Fran\c cois-Rabelais Tours,\\
Parc de Grandmont, 37200 Tours, France}                                         

\vspace{3pt}{\hoch{4}\it 
LE STUDIUM, Loire Valley Institute for Advanced Studies,\\
Tours and Orleans, France}

\vspace{3pt}{\hoch{5}\it George P. \& Cynthia W. Mitchell
Institute for\\ 
Fundamental Physics and Astronomy,\\ Texas A\&M University,
College Station, TX 77843-4242, USA}

\vspace{3pt}{\hoch{6}\it 
 Center for Applied Mathematics and Theoretical Physics,
 University of Maribor, SI2000 Maribor, Slovenia}
\vspace{3pt}

\begin{abstract}
{\textnormal{
We present three families of exact, cohomogeneity-one Einstein metrics 
in $(2n+2)$ dimensions, which are generalizations of the Stenzel 
construction of Ricci-flat metrics to those with a positive cosmological 
constant.  The first family of solutions  are Fubini-Study metrics on 
the  complex projective spaces $\CP^{n+1}$, written in a Stenzel form, 
whose principal orbits 
 are the Stiefel  manifolds $V_2(\R ^{n+2})=SO(n+2)/SO(n)$ divided by $\Z_2$.  
The second family are also Einstein-K\"ahler metrics, now on the 
Grassmannian manifolds $G_2(\R^{n+3})=SO(n+3)/((SO(n+1)\times
SO(2))$, whose principal orbits  are the Stiefel manifolds 
$V_2(\R^{n+2})$ (with no $\Z_2$ factoring in this case).  
The third family are  Einstein metrics on the product 
manifolds $S^{n+1}\times S^{n+1}$, and are K\"ahler only for $n=1$. 
Some of these metrics are believed to play a  role in studies
of consistent string theory compactifications and 
in the context of the AdS/CFT correspondence. We also elaborate on 
the geometric approach to quantum mechanics based on the K\"ahler 
geometry of  Fubini-Study metrics on  $\CP^{n+1}$, and we apply the
formalism to study the quantum entanglement of qubits. }}

\end{abstract}

\end{center}

\pagebreak


\tableofcontents
\addtocontents{toc}{\protect\setcounter{tocdepth}{2}}

\section{Introduction}

The study of cohomogeneity-one Einstein  metrics
by employing  the techniques used in homogeneous 
cosmology \cite{Ryan:1975jw} was initiated 
in \cite{Belinsky:1978ue,Gibbons:1978zy,Gibbons:1979xn,Gibbons:1989er}.
The  Einstein equations lead to second-order differential
equations which were shown to follow  from  a suitable Lagrangian. 
Imposing the condition that the metric have reduced holonomy 
was shown  to  lead  to first-order differential equations
that implied the second-order equations. 
In many cases these first-order equations admit simple explicit
solutions. It was later shown
that  in many  cases when this reduction is  possible, the potential
 may be derived from a superpotential \cite{cvgilupo}. A particularly
interesting class of examples consists of $(2n+2)$-dimensional 
metrics with the isometry group $SO(n+2)$,
and these are 
the subject of the present paper. Specifically, the metrics we shall
consider have cohomogeneity one, with level surfaces that are
homogeneous squashed Stiefel manifolds 
$V_2(\R ^{n+2}) \equiv  
O(n+2)/O(n) \equiv SO(n+2)/SO(n)$, consisting of the set
of orthonormal dyads in $\R^{n+2}$  \cite{Schwarz}
\footnote{The reader is warned that there appears to be no standard
 notation for the  Stiefel manifolds  $V_p(\mathbb{R}^n)$ and their  
cousins the Grassmannian manifolds  $G_p(\mathbb{R}^n)$.
For us and  in \cite{Schwarz}, the Stiefel manifold 
\break $V_p(\mathbb{R}^{p+q})= O(p+q)/O(p) = SO(p+q)/SO(p)$  is the space 
of  $p$-frames in $\R^{p+q}$.  However, we differ  from \cite{Schwarz} 
on Grassmannian manifolds. 
For us  $G_p(\mathbb{R}^{p+q})=SO(p+q)/(SO(p) \times SO(q))$ is the space of
\emph{oriented} $p$-planes in $\R^{p+q}$.  In \cite{Schwarz}  
$G_p(\mathbb{R}^{p+q})= O(p+q)/(O(p)\times O(q))$ is the space of 
\emph{un-oriented}
$p$-planes in $\R^{p+q}$. The latter is a $\mathbb{Z}_2$ quotient of 
the former.}. 
In addition to the 
references cited above, some relevant  previous  work
can be  found  in  
\cite{Pedersen, bougib,stenzel,Dancer:1993aw, Dancer:2002cq,Cvetic:2002kj}.

Perhaps the best-known example of a  metric in the class we shall be 
considering is Stenzel's  Ricci-flat 6-metric on the tangent bundle of 
the 3-sphere
\cite{stenzel}, which figures in string theory as the 
deformed conifold \cite{Candelas:1989js}. Recently,
Kuperstein \cite{kuperstein} has studied 
the behaviour of the conifold in the presence
of a positive  cosmological constant, and he found
numerical evidence  for a solution  of a set of 
first-order equations that provides  
a complete  non-singular cohomogeneity-one Einstein metric on a  
``compactification''
of $T^\star S^3$. The 6-manifold is fibred by an open interval of 
five-dimensional  principal orbits which  degenerate at 
 one end  of the interval to an $S^3$ orbit, as in the case
of the deformed conifold, and at the other end to an $S^2 \times S^2$   
orbit.

 In this paper, we construct three families of simple exact solutions to the
equations of motion for Stenzel-type Einstein metrics with a positive
cosmological constant, and we study the global structures of the 
manifolds onto which these local metrics extend.  Although the
metrics are written in a cohomogeneity-one form, all three classes of
metrics that we obtain are actually homogeneous.  The first class of
solutions we obtain, which satisfy the first-order equations and therefore
are Einstein-K\"ahler, extend smoothly onto the manifolds of the 
complex projective spaces $\CP^{n+1}$.  In fact, as we subsequently 
demonstrate, these are precisely the standard Fubini-Study metrics on
$\CP^{n+1}$, but written in a rather unusual form.  The principal orbits 
of these metrics are the Stiefel   manifolds $V_2(\R ^{n+2})$, divided by
$\Z_2$.
The $\CP^{n+1}$ manifold is
described  in a form where there is an $S^{n+1}$ degenerate orbit or  
bolt at one end of the range of the
cohomogeneity-one coordinate, and
an $SO(n+2)/(SO(n)\times SO(2))/\Z_2$ bolt at the other end.
The case $n=1$, giving $\CP^2$, corresponds to a  solution  of the
first-order equations obtained by a  geometrical construction presented
in \cite{bougib}.  Here, we give  a generalization of  this
construction to all values of $n$. 

   We find also a second family of exact solutions of the first-order
equations.  We demonstrate that these Einstein-K\"ahler metrics extend
smoothly onto the Grassmannian manifolds $G_2(\R^{n+3})=SO(n+3)/((SO(n+1)\times
SO(2))$  of {\it oriented} 2-planes
in $\R^{n+3}$. 
The level surfaces are again the Stiefel manifolds 
$V_2(\R^{n+2}) \equiv SO(n+2)/SO(n)$, which can 
be viewed as $U(1)$ bundles over the Grassmannian 
manifolds  $G_2(\R^{n+2})=SO(n+2)/(SO(n)\times
SO(2))$.  (In these metrics, unlike the $\CP^{n+1}$ metrics 
described above, the Stiefel manifolds of the principal orbits are
not factored by $\Z_2$.)  
The metric we obtain on $G_2(\R^{n+3})$  is homogeneous,
described as a foliation of squashed Stiefel manifolds 
$V_2(\R^{n+2})=SO(n+2)/SO(n)$.  The metric has 
an $S^{n+1}$ bolt at at one end of the range of the
cohomogeneity-one coordinate, just as  in the Stenzel form of the 
$\CP^{n+1}$ metric, and an
 $SO(n+2)/(SO(n)\times SO(2))$ bolt at the other end.
The case $n=2$,
corresponding to the Grassmannian $G_2(\R^5)$, is in fact 
the exact solution for an Einstein-K\"ahler metric that was found 
numerically by Kuperstein in \cite{kuperstein}.  

   The third family of metrics that we obtain arises as solutions
of the second-order Einstein equations, but they do not, in 
general, satisfy the
first-order equations. Thus they are Einstein but not K\"ahler.
We provide a geometrical 
construction for those metrics, which demonstrates that they
extend smoothly onto the product manifolds $S^{n+1}\times S^{n+1}$.
In the case $n=1$, the geometrical 
construction coincides with one
first given in \cite{Pope} and described  in detail
in appendix B of \cite{Cvetic:2002kj}.  The $n=1$ case is exceptional in that
the metric, on $S^2\times S^2$, is K\"ahler as well as Einstein.

Some of the metrics discussed in this paper may play a role in studies
of consistent M-theory or string theory compactifications, and 
in the context of the
AdS/CFT correspondence. For example, a consistent compactification of
Type IIA supergravity on $\CP^{3}$ results in an $N=6$ supersymmetric
four-dimensional gauged supergravity theory.  This was shown in
\cite{nilpop}, where it was obtained via a reduction of the $S^7$ 
compactification of $D=11$ supergravity on the Hopf fibres of the $S^7$
viewed as a $U(1)$ bundle over $\CP^3$.
The $\CP^{n+1}$ spaces also provide a natural base for constructions of
elliptically fibered Calabi-Yau $(n+2)$-folds, relevant to studies of
F-theory compactifications to $(8-2n)$ dimensions 
(c.f. \cite{CveticKlevers} and references therein).  In the context of
the AdS/CFT correspondence,  $\CP^{n+1}$ or $G_2(\R^{n+3})$ backgrounds, as opposed to compact
Calabi-Yau $(n+1)$-folds, have the possibility of avoiding the appearance of
singular D-$(p+2)$ brane fluxes in the presence of anti-D-$p$ branes (c.f.
\cite{Gautason} and references therein).

The relevance  of the metrics discussed in this paper
is not only restricted   to problems in quantum gravity and in M-theory or
string theory.
The ideas presented in appendix B of  \cite{Cvetic:2002kj} were taken from the
quantum theory of triatomic molecules in the Born-Oppenheimer approximation. 
At a more fundamental level, $\CP^{n}$ is the space of physically distinct 
quantum states of a system with an $(n+1)$-dimensional Hilbert space, and  
forms the arena  for  the geometrical approach to quantum mechanics that
exploits the K\"ahler geometry of its Fubini-Study metric
 \cite{Kibble,Gibbons:1991sa,Brody:1999cw}. The calculations
on $\CP^2$ 
in  \cite{bougib}  were aimed at  evaluating 
the   Aharonov-Anandan phase for a 3-state
spin-1  system using  the K\"ahler connection. 
More recently there have been interesting 
applications using ideas from toric geometry \cite{Aadel:2015jda}.  
In this paper we shall further elaborate on applications of this formalism,
including the study of the quantum entanglement of qubits.

     The paper is organised as follows. In section 2 we give a brief 
outline of the geometric approach to quantum mechanics, and its further 
applications.  This includes a discussion of the quantum entanglement of
systems comprising two qubits and three qubits. 
In section 3 we  summarise the Stenzel construction of 
the Ricci-flat metrics on the tangent bundle of $S^{n+1}$, which lends 
itself to the generalisation that allows us to construct Einstein-K\"ahler 
metrics with a positive cosmological constant.  In section 4 we  
construct the explicit Einstein-K\"ahler metrics of the Stenzel type 
on $\CP^{n+1}$, and analyse  their global structure. In Section 5 we 
obtain the Einstein-K\"ahler metrics  on the Grassmannian manifolds 
$G_2(\R^{n+3})=SO(n+3)/(SO(n+1)\times SO(2))$, as further 
exact solutions of the first-order equations for the metrics of Stenzel
type.  We also obtain exact solutions of the second-order equations,
for Einstein metrics of the Stenzel type 
that are not, in general, K\"ahler, on the
product manifolds $S^{n+1}\times S^{n+1}$.  Furthermore, by means of
analytic continuations we obtain the 
metrics, with negative cosmological constant, on the 
non-compact forms of the $\CP^{n+1}$, $G_2(\R^{n+3})$ 
and $S^{n+1}\times S^{n+1}$ manifolds. 
In section 6 we discuss the case of six dimensions in detail, with an
explicit coordinatisation of the left-invariant 1-forms on the
five-dimensional principal orbits.  We also provide a detailed comparison of our exact solutions  with Kuperstein's numerical and asymptotic analysis. 
A summary and conclusions  are given in Section 7.

\section{Quantum Mechanics on $\CP ^{n}$} 

The goal of this section it to spell out the key steps in  
formulating a geometric approach to quantum mechanics, based on the 
K\"ahler geometry of the Fubini-Study metric on $\CP^{n}$.
We begin by reminding the reader that in the standard formulation
of quantum mechanics, Schr\"odinger's equation
is just a special case of Hamilton's equations \cite{Dirac,Strocchi}.
Let  $|a \rangle$, for $a=1,2 \dots,n+1$, 
be an orthonormal  basis for $ \C^{n+1}$,
and 
\ben
|\Psi\rangle = Z^a\,  |a \rangle \,, \quad Z_a = \frac{1}{\sqrt 2}
 (q^a+ i p_a) \,,\quad H(q^a,p_a, t)= 
\langle 
\Psi| \hat H | \Psi  \rangle  = \bar Z^a H_{ab} Z^b\,,
 \een
where  $q^a \in \R^{n+1}$, $p_a \in \R^{n+1}$ and 
$H_{ab}= \langle a|\hat H |b \rangle 
= \bar H_{ba} $. Thus  
\ben
\frac{d Z^a} {d t}= \frac{1}{i} \frac{\p H}{\p  \bar Z ^a}\,, 
 \een
or 
\ben
\frac{d q^a}{dt} = \frac{\p H}{\p p_a} \,,\qquad  \frac{d p_a}{dt}
= -\frac{\p H}{\p q^a } \,.  
 \een
In effect, we are making use of the fact that $\C ^{n+1}$, 
considered as a Hilbert space, is a flat K\"ahler manifold  with
K\"ahler potential  $K=\bar Z^a Z^a $, metric 
\ben
ds^2 = 
\big| d|\Psi \rangle\big|^2 =  \frac{\p ^2 K}{ \p Z^a \p {\bar Z}^a }   
   d \bar Z ^a d Z^a \, = d \bar Z^a dZ^a   = \half 
(d q ^a dq^a + dp_a dp_a) \,, 
\een
symplectic form
\ben 
\omega =  \frac{1}{i} \,  \frac{\p ^2 K}{ \p Z^m \p {\bar Z} ^n }
\,  dZ^ m \wedge d { \bar Z} ^n =   \frac{1}{i} d \bar Z^a \wedge d  \bar Z ^a 
  = dp_a \wedge d q^a  \,,  
\een 
and complex structure
\ben
J \frac{d q^a}{d t} = \frac{d p_a}{d t} \,,\quad  
J \frac{d p_a }{d t} = - \frac{d q^a }{  dt} \,.  
\een

   This formalism, however, has a built-in
redundancy, since  $|\Psi \rangle$ and $\lambda |\Psi \rangle$
 with $\lambda$ a non-vanishing complex number are physically  
equivalent states.
We can partially fix this freedom by normalising our states, requiring that
\ben
\langle \Psi | \Psi \rangle = \bar Z ^a Z^a =1 \,.
\een
This restricts the states to  $S^{2n+1} \subset \R^{2n}$,   
but it still leaves the freedom to change the overall phase:
$|\Psi \rangle  \rightarrow  e^{i \alpha} \,  |\Psi \rangle$
with $\alpha \in \R$. To obtain the space of 
physically-distinct states,  we must therefore 
take the quotient $S^{2n+1}/U(1)$. 
As a  complex manifold this is just  $\CP^n$, with the orbits
of the $U(1)$ action being the Hopf fibres.
An atlas of complex coordinates is provided by the 
inhomogeneous coordinates
$\zeta ^a_b = Z^a/Z^b$, $a \ne  b$.   
 
   In order to endow
 $\CP^n$ with a metric, we project the standard round metric on $S^{2n+1}$ 
orthogonally to the fibres:
\ben
ds^2  = \big|d | \Psi \rangle\big|^2 -| \langle \Psi | d | \Psi \rangle   |^2   
= d\bar Z ^ a d Z^a - | \bar Z^a d Z^a |^2 \,. \label{FBmetric}   
\een
Introducing the inhomogeneous coordinates $\zeta ^i = Z^i/Z^{n+1}$ ,
 $i=1,2,\dots,n$ we find that the K\"ahler form is  
given by
\ben
K= \log (1+ \bar \zeta^i \zeta^i ) \,.
\een
If $n=1$ we get the Bloch sphere \cite{Bloch}, with 
metric  $\frac{1}{4}$ the unit round  metric on
$S^2$. This is the space of spin $\half$ states, 
or of a single qubit. For a spin-$J$ state we get 
$\CP^{2J}$. If $J=1$ one speaks of a q-trit and in general
a  q-dit with $d=(2J+1)$. For $N$ qubits we have $n+1=d=2^{N}$,
because  in this case  the Hilbert space is $(\C^2)^{\otimes N} $
and not $(S^2)^N$ as one might imagine for $N$ classical spin-$\half$ 
particles.

   The physical significance of the Fubini-Study metric is that
the distance  $s^{\phantom\Sigma}_{FB}$ between  two states 
$|\Psi\rangle$ and $|\Psi^\prime\rangle$  
is given in terms of the transition probability 
$|\langle \Psi| \Psi ^\prime \rangle|^2$ between
the two states by  
\ben
\cos ^2 (s^{\phantom\Sigma}_{FB}) = 
|\langle \Psi| \Psi ^\prime \rangle|^2 \,.\label{trans}\een
Since in inhomogeneous coordinates 
\ben
|\Psi \rangle = \frac{1}{\sqrt{1+|\zeta|^2}} \Bigl (\zeta^i \, |i \rangle 
+ |n+1\rangle \Bigr )\,,
\een
we have 
\ben
\cos  (s^{\phantom\Sigma}_{FB}) =  \frac {| 1 + \bar \zeta^i \zeta^i  |}
 {\sqrt{ (1+|\zeta|^2 )(1+ |\zeta ^\prime|^2)} }  \,.        
\een

   The instantaneous velocity of the evolution of a normalised state 
$|\Psi  \rangle$ under the action
of a Hamiltonian $\hat H$, which could be time-dependent
is,  using (\ref{FBmetric}), given by 
\ben
\frac{d s^{\phantom\Sigma}_{FB}}{dt} = \sqrt{
 \langle \Psi| \hat H^2 | \Psi \rangle  
-  \bigl ( \langle \Psi| \hat H | 
 \Psi \rangle  \bigr)^2      }  = \Delta E \,, \label{speed} 
\een   
where $\Delta E$ is the instantaneous root mean square deviation of
the energy in the state  $|\Psi  \rangle$. Note that (\ref{trans}) and 
(\ref{speed})  are discrepant  by a factor of two 
from \cite{Anandan:1990fq}, whose metric is 4 times the 
Fubini-Study metric, that is $s^{\phantom\Sigma}_{AA}=
 2 s^{\phantom\Sigma}_{FB}$.

\subsection{Darboux coordinates and shape space}

One may replace  the inhomogeneous coordinates
  $\zeta^i$   by  
\ben
a^i=\frac{\zeta^i}{\sqrt{1+|\zeta|^2 }} \,, \qquad \Longleftrightarrow  
\qquad \zeta^i= \frac{a^i}{\sqrt{1-|a|^2 }} \,,
\een 
by which is an open dense subset of $\CP^n$ 
is mapped into the interior
of the unit ball in ${\C} ^n \equiv{\R} ^{2n}$.
Since if $K= \log(1+|\zeta|^2) $, 
\ben
\half \frac{\p^2 K}{\p \zeta ^ i \p \bar \zeta ^j} d \zeta ^i \wedge d \bar \zeta ^j
= \half d a^i \wedge d \bar a ^i = i dp_i \wedge dq^i    
\een
where   $a^k= q^k + i \,p_k$ .
Thus  $(q^i,p_i) $ are Darboux coordinates 
for ${\CP} ^n $. If $n=1$ we recover what geographers
call the coordinates associated to {\it Lambert's Polar Azimuthal
Equal Area Projection}. By contrast, if $n=1$ and the  
inhomogeneous coordinate
$\zeta^1$ is  used,  we have what astronomers
and crystallographers know as  {\it 
The Equal Angle Stereographic Projection of Hipparchus}. 

In terms of the Lambert-Darboux coordinates we have 
\ben
|\Psi \rangle =  a ^i |i \rangle 
+  \sqrt{1- |a|^2} \, |n+1\rangle \,,
\een
and hence $H= \langle \Psi |\hat H |\Psi \rangle$ is given by 
\ben  H
=  {\bar a} ^i H_{ij} a^j +  
(1- |a|^2 ) H _{(n+1) \,(n+1) }
+ \sqrt{1-|a|^2 } \bigl( {\bar a}^i\, H_{i\,(n+1)}  + 
H_{(n+1)\,i} \, a^i \bigr ) \,,
\een
which is considerably simpler than its expression
in inhomogeneous coordinates 
\bea
H&=& \frac{1}{(1+ |\zeta|^2 ) } \Bigl( {\bar \zeta} ^i H_{ij} \zeta ^j  
+  {\bar \zeta }^i H_{i\,(n+1)}  + 
H_{(n+1)\,i} \, \zeta ^i   +  H _{(n+1) \,(n+1) } \Bigr )    
 \,,
\eea
In particular,  if $  H_{(n+1)\,i} =0$ , the Hamiltonian is purely
quadratic in the Lambert-Darboux coordinates. It is possible to
express $\cos \delta_{FS}$ and the Fubini-Study metric
in terms of Lambert-Darboux coordinates, but
the expressions don't appear  to be especially illuminating.
  
There is interesting application  of the foregoing theory
to the statistical theory of shape \cite{Kendall1,Kendall2,Brodyshape}. 
A shape is defined to be a set of $k$ labelled   points
$\bx_a$, $a=1,2,\dots n$  in $\R^n$ modulo the action of the similarity 
group Sim$(n)$, i.e the group of
translations, rotations and dilations. The space of such shapes is 
denoted by $\Sigma_n^k$ and hence has dimension $nk -n  - \half n(n-1)-1$.
If we translate  the $k$ points  so that their centroid lies at  the origin
of $\R^n$, and we fix the scale by demanding that
\ben
 \sum_1^{k-1} \bx_i^2 =1\,,
\een    
we see that 
\ben
\Sigma_n^k= S^{n(k-1) -1} /SO(n) \,.
\een
Moreover, the flat metric on $\R^{n(k-1)}$  descends to give a curved metric
on $\Sigma _n^k$. 

   In the special case when $n=2$, we find
that $\Sigma_2^k  = S^{2k-3}/SO(2) =   \CP ^{k-2}$,  
with its Fubini-Study metric. Thus 
the space of triangles in the plane may be identified with the Bloch sphere
$\CP^1$. Using complex notation, the $k-1$ coordinates $Z^i$, may be regarded
as  homogeneous coordinates for $ \CP^{k-2}$. The inhomogeneous coordinates
are  $\zeta^i= Z^i/Z^{k-1} $, $\ i=1,2,\dots, k-2$, and the Darboux coordinates
are 
\ben
a^i=  \frac{Z^i}{Z^{k-1}} \frac{1}
{\sqrt{1+ \sum_j^{k-2} |Z^j|^2/|Z^{k-1}|^2   }} = 
e^{-i\theta_{k-1}} Z^i\,,\qquad i=1,2,\dots, k-2\,,    
\een    
where $\theta_{k-1}$ is the argument of $Z^{k-1}$. Thus if 
$e ^{-i\theta_{k-1}} Z^i =  x ^i +i y^i $, the volume measure on the 
shape space $\Sigma_2^k$ 
is uniform in these Lambert--Darboux coordinates, i.e. it is 
\ben
 \prod_1^{k-2}  dx^i dy^i \,.
\een 
For a description  of entanglement and other aspects of quantum mechanics in 
terms of shapes see \cite{Brodyshape}.

\subsection{Entanglement and Segre embedding}
 
As noted above,
the Hilbert space for two qubits
is $\C^2 \otimes \C^2=\C^4$, and the space of states  is $\CP^3$, which as 
a real manifold is six dimensional. However, for two non-interacting 
completely independent spin-half  systems, each of  whose state spaces is the 
Bloch sphere $\CP^1 =S^2$, one might expect a state space of the form 
$\CP^1\otimes \CP^1= S^2 \times S^2$. This will be the case if   
we  consider only  separable  or unentangled states   
in $\C ^4 = \C^2 \otimes \C^2 $, for which     
\ben
\ket = \ket _1 \otimes \ket _2\,, 
\een
with 
\ben
\ket _1 = a_1 \uket_1  + b_1 \dket_1  \,,\qquad \ket _2 = 
a_2 \uket_2 + b_2 \dket_2 \,.
\een
If $\uuket = \uket_1 \otimes \uket_2$, etc., then 
\ben
\ket = Z^1 \uuket + Z^2 \udket + Z^3 \duket + Z^4 \ddket \,, 
\een
with \ben
(Z^1\,,Z^2\,, Z^3\,, Z^4) = (a_1a_2 \,, a_1b_2\,, b_1 a_2 \,, b_1 b_2 ) \,,
\een 
and so there is a non-linear constraint
on the set of bi-partite states, namely
\ben
Z^1 Z^4 = Z^2 Z^3  \,. \label{hyper}
\een
We conclude that  the set of all  separable states with respect to this
factorization of the Hilbert space $\C^4$  is not a 
linear subspace 
of $\C^4$, but rather (\ref{hyper}) is  a complex quadratic cone in 
$\C^4$. This
 projects down to a complex hypersurface  in  $\CP ^3$, 
given, in terms of the inhomogeneous coordinates 
$(\zeta ^1\,,\zeta ^2 \,, \zeta ^3) =
(Z^1/Z^4\,,Z^2/Z^4\,,Z^3/Z^4)$, by
\ben
\zeta^1 = \zeta ^2 \,\zeta ^3\,.
\een 
The K\"ahler function for  $\CP^3$ is 
\ben
K= \log(1+ |\zeta ^1|^2 + |\zeta ^2|^2 +  |\zeta ^3|^2 )
\een
and so this restricts to
\ben
K= \log(1+ |\zeta ^2 \zeta ^3 |^2 + |\zeta ^2|^2 +  |\zeta ^3|^2)
= \log (1 + |\zeta ^2|^2  ) + \log ( 1 + |\zeta ^3|^2   ) \,.  
\een
Thus we get the product of Fubini-Study metrics on $\CP^1 \times \CP^1$.
This construction and its generalizations are known to mathematicians 
as Segre embeddings. 
In physical terms, a linear superposition of unentangled states 
is, in general, entangled. The span of all such states, 
that is, the union of  complex lines on $\CP^3$  through
all pairs of points on  the  Segre embedding
of $\CP^1 \times \CP^1$ into $\CP^3$,  is all of $\CP^3$. 

   The simplest notion of entanglement depends upon the
factorization of the total Hilbert space into
a tensor product of two  Hilbert spaces.
In our present case, since $2\times 2=2+2$ the second
Hilbert space is orthogonal,
\ben
\cH=\cH_\otimes\cH_s = \cH_1 \oplus_\perp \cH_2 \,.
\een 
Each factorization amounts to finding a two-dimensional linear subspace
of $\C^4$.   The space of such linear subspaces is  
is the complex  Grassmannian  $G_2( \C^2) = SU(4)/(SU(2)_1\times SU(2)_2)$,
where $SU(2)_1$  acts on $\cH_1$ and $SU(2)_2$  acts on $\cH_2$.
In fact this is the only such simple case since the only integral solution
of the equation $n_1 n_2= n_1+n_2 $ is $n_1=n_2=2$.
 
    One  physical situation where this decomposition arises
is when $SU(2)_1$ is isospin and $SU(2)_2$ is ordinary spin.
Then  $\uket \otimes |\Psi \rangle_2 $ are states of the 
the proton  with electric  charge $|e|$ and 
$\dket \otimes |\Psi\rangle_2$ are states of the   
neutron with zero electric charge \cite{Heisenberg}. 
Since electric charge is absolutely conserved,
we have a super-selection rule \cite{WWW}; no other superpositions are allowed.
Thus the proton states correspond to a point at the north pole
of $S^2_1\times S^2_2$ and the neutron states to a point at the south
pole of $S^2_1\times S^2_2$. 

\subsection{Tripartite entanglement and Cayley hyperdeterminant}

The  significantly more complicated case of three qubits with 
the possibility of tripartite
entanglement 
\ben
\C^8 = \C^2 \otimes \C^2 \otimes \C^2 \,,
\een
which may be quantified  by means  the Cayley hyperdeterminant 
\cite{cayley,gelkapzel},
has arisen recently \cite{duffrev,Borsten:2012fx}
in the study of STU  black holes \cite{cvety,cvett}.
If we adopt a  binary digit notation, according to which 
$\uparrow$ corresponds to 0
and $\downarrow$ corresponds to 1,    we have 
\bea
&&\bigl( \zeta^1 \, |0\rangle_1 + |1 \rangle_1 \bigr ) \otimes 
\bigl( \zeta^2\, |0\rangle_2 + |1 \rangle_2 \bigr ) 
\otimes \bigl( \zeta^3\, |0\rangle_3 + |1 \rangle_3 \bigr ) \nonumber\\
 &=&\zeta^{000}\, |000\rangle + \zeta ^{001}\, |001 \rangle 
+ \zeta^{100} \, |100\rangle  \nonumber \\
&+& \zeta ^{010}\, |010\rangle + \zeta^{110}\, |110\rangle 
  + \zeta ^{101}\, |101 \rangle
+ \zeta ^{011}\,  |011 \rangle + |111\rangle \,, \label{product}
 \eea
where $(\zeta ^1,\zeta ^2, \zeta ^3)$ are inhomogeneous coordinates for 
$\CP^1 \times \CP^1\times \CP^1$  and $(\zeta ^{000}, \dots, \zeta ^{011} )$
are inhomogeneous coordinates for $\CP^7$. 
In this case  the Segre embedding is given (locally) by 
\bea 
(\zeta^{011},\zeta ^{101}, \zeta ^{110} )   &=& 
( \zeta ^1, \zeta ^2 ,\zeta ^3 )\,, \nonumber \\
(\zeta ^{001}, \zeta ^{100}, \zeta ^{010}) &=& 
(\zeta ^1 \zeta ^2,\zeta ^2 \zeta ^3,\zeta ^3 \zeta ^1 )  \,,
\nonumber \\
\zeta ^{000} &=& \zeta ^1 \zeta ^2 \zeta ^3 \,.
\eea
or as a sub-variety of $\CP^7$ by  the four equations in seven unknowns
\bea
(\zeta^{001},\zeta ^{010}, \zeta ^{100} )  &=& ( \zeta^ {011}\, \zeta^{101},
\zeta ^{011}\, \zeta ^{110}, \zeta ^{110} \, \zeta^{101})\,,\nn\\   
\zeta ^{000} &=& \zeta ^{011}\, \zeta ^{101} \, \zeta ^{110} 
\,. \label{Segre}
\eea

In \cite{Borsten:2012fx} the general state in $\C^8$ is written as
\ben
|\psi \rangle = \sum_a \psi_a |a \rangle \,,
\een  
where $a = 0,1,\dots,7$ correspond to the binary digits used above.
Thus
\ben
(\psi_0, \psi_1,\psi_2,\psi_3,\psi_4,\psi_5,\psi_6,\psi_7)=
 (\zeta^{000}, \zeta^{001}, \zeta ^{010},\zeta^{011},\zeta^{100},\zeta^{101},
\zeta ^{110} ,1)  \,. \label{psi}
\een 
The Cayley hyperdeterminant is given by \cite{cayley,gelkapzel}
\be
D(\zeta) = -\ft12 b^{ij}\, b^{k\ell}\, \ep_{ik}\, \ep_{j\ell}\,,\qquad
\hbox{where}\qquad b^{ij}=\zeta^{ik\ell}\, \zeta^{jmn}\, \ep_{km}\, 
\ep_{\ell n}
\ee
and $\ep_{ij}=-\ep_{ji}$ with $\ep_{01}=1$.  In terms of the components 
$\psi_a$, this implies
\bea
D(|\psi\rangle)&=&  \Bigl( \psi_0\psi_7 -  \psi_1\psi_6 -    
\psi_2\psi_5 -   \psi_3\psi_4    \Bigr )^2  
   -4 \Bigl( \psi_1 \psi_6 \psi_2 \psi _5  
+ \psi_2 \psi_5 \psi_3 \psi _4
+  \psi_3 \psi_4 \psi_1 \psi _6 \Bigr )\nonumber \\
&+& 4 \psi_1 \psi_2 \psi_4  \psi_7  + 4 \psi_0 \psi_3 \psi_5 \psi_6 
\,. \label{tangle}\eea 
Substituting in (\ref{Segre}), we see that 
the Cayley hyperdeterminant or {\it three-tangle} vanishes on the image
of the Segre embedding, as expected.  We can also see the embedding
geometrically, in that the K\"ahler function for $\CP^7$,
\be
K_7= \log(1+  |\zeta^{000}|^2 + |\zeta^{001}|^2 + 
   |\zeta^{010}|^2 + |\zeta^{011}|^2 +  |\zeta^{100}|^2 + |\zeta^{101}|^2 +
    |\zeta^{110}|^2)\,,\label{K7function}
\ee
becomes the sum of K\"ahler functions for three $\CP^1$ factors after using
the equations (\ref{Segre}):
\be
K_7\longrightarrow \log(1+|\zeta^1|^2) + \log(1+|\zeta^2|^2) +
               \log(1+|\zeta^3|^2)\,.
\ee
If the components $\psi_a$ are taken to be real, then the entropy of the 
BPS STU black holes \cite{cvety,cvett}  and the  Cayley hyperdeterminant 
are related  by \cite{duffrev,Borsten:2012fx}: 
\be
S=\pi\sqrt{-D(|\psi\rangle)}\, , 
\ee
provided that the four electric $\{q_i\}$  and four  magnetic$\{p^i\}$  
charges  are identified as: 
\be
(p^0, p^1, p^2,p^3,q_0,q_1,q_2,q_3)=(\psi_0,\psi_1,\psi_2,\psi_4,-\psi_7,\psi_6,\psi_5,\psi_3)\, . 
\ee

$\C^8$ also admits a  bi-partition
as $\C^2 \times \C^4$, and thus a Segre embedding of $\CP^1 \times \CP^3$. 
This works out as follows. The analogue of (\ref{product}) is  
\bea
\bigl( \zeta^0 \, |0\rangle _1 + |1\rangle _1\, \bigr ) &\otimes& 
\Bigl(\zeta^1 \,|0\rangle_2 \otimes   |0\rangle_3  \nonumber +
\zeta ^2 \, |1\rangle_2 \otimes  |0\rangle_3 +\zeta ^3 \, 
|0\rangle_2 \otimes  |1\rangle_3 +  |1\rangle_2 \otimes  |1\rangle_3 
\Bigr )\\ & =& \sum_a \psi_a  |a \rangle \,. 
\eea
The analogue of (\ref{Segre}) is  
\ben
(\psi_0, \psi_1,\psi_2,\psi_3,\psi_4,\psi_5,\psi_6,\psi_7)=
(\zeta^0 \,\zeta^1, \zeta^0\, \zeta^3, \zeta^0\,\zeta^2, \zeta^0, 
 \zeta^1, \zeta^3, 
\zeta^2,1) \,,  \label{sip}
\een
giving three equations in seven unknowns:
\ben
(\psi_0,\psi_1,\psi_2)= (\psi_3 \psi_4,\psi_3 \psi_5,\psi _3 \psi_6 )\,,
\label{Segre13}
\een  
or in other words
\be
\zeta^{000}= \zeta^{011}\, \zeta^{100}\,,\qquad
\zeta^{001}=  \zeta^{011}\, \zeta^{101}\,,\qquad
\zeta^{010}= \zeta^{011}\, \zeta^{110}\,.
\ee
Substitution of (\ref{sip}) in (\ref{tangle}) shows that
the Cayley hyperdeterminant of the three-tangle vanishes in this case as well.
We also find that the K\"ahler function (\ref{K7function}) for $\CP^7$
becomes the sum of K\"ahler functions for a $\CP^1$ and a $\CP^3$ factor
after imposing the conditions (\ref{Segre13}):
\be
K_7 \longrightarrow \log(1+|\zeta^0|^2) +
\log(1+|\zeta^1|^2 + |\zeta^2|^2 + |\zeta^3|^2)\,.
\ee

\subsection{Direct sums and nesting formulae}

We have seen above that as  well as partitions into tensor products, 
it is often convenient  to
decompose Hilbert spaces into direct sums. This gives rise
to an iterative   ``nesting construction'' 
for Fubini-Study metrics \cite{Hoxha:2000jf}. 

Consider the case 
\ben
\C^{p+q} = \C^p \oplus_\perp  \C^q
\een
with $p\ge q$. Let  
\ben
Z= \begin{pmatrix} \cos \alpha \, X \\ \sin \alpha\,  Y \end{pmatrix}  \,,
\een
with 
\ben
X^\dagger X= Y^\dagger Y = 1 \,,
\een
and hence $Z$ is a unit vector in $\C^{p+q}$:
\ben
Z^\dagger Z =1 \,.
\een
If  we define $d \Sigma^2_{m}$, to be  the Fubini-Study  metric  
(\ref{FBmetric})
on $\CP^{m}$, we have  
\crampest
\bea
d \Sigma^2_{p+q-1} &=&    dZ^\dagger dZ -| Z^\dagger dZ|^2 \nonumber\\ 
&=& d \alpha ^2 + \cos^2 \alpha \,\bigl(  dX^\dagger dX -| X^\dagger dX|^2              \bigr )  + \sin ^2 \alpha 
\, \bigl(  dY^\dagger dY -| Y^\dagger dY|^2 \bigr) \nonumber \\
&+& \cos ^2 \alpha \sin ^2 \alpha \,|X^\dagger d X + Y^\dagger dY |^2 \\
&=& d\alpha ^2  + \cos^2 \alpha \,d\Sigma_{p-1}^2 + 
\sin^2 \alpha \,d\Sigma^2 _{q-1}  + 
\sin^2 \alpha \cos ^2 \alpha  \,|X^\dagger dX + Y^\dagger  dY|^2 \,,
\label{nest}
\eea 
\uncramp
where  we have have used the fact that
\ben
\Re \,X^\dagger dX = \Re \,Y^\dagger dY =0 \,.
\een
Note that $-\im X^\dagger dX$ and $-\im Y^\dagger dY$ are the
K\"ahler connections on $\CP ^{\,p-1}$ and $\CP^{\,q-1}$ respectively.

If $p=n$, $q=1$, $Y= e^{i\bar \tau}$ and $\alpha =\ft12 \pi - \xi$ , 
we recover the  iterative  construction of   \cite{Hoxha:2000jf},
in which given the Fubini-Study metric on $\CP^n$, one obtains
the Fubini-Study metric on $\CP^{n+1}$. Carrying out the
iteration gives the metric as a nested sequence of metrics
ending with the the round metric on $\CP^1$. In the first non-trivial
case, one obtains $\CP^2$  in Bianchi-IX form  \cite{Gibbons:1978zy}. 
It is clear that one may decompose the higher-dimensional
metrics into further  direct sums by using (\ref{nest})
applied to $d\Sigma^2 _{q-1}$  or $ d\Sigma^2 _{q-1}$ or both. 

\section{The Stenzel Construction}

  We begin by recalling the Stenzel construction of $(2n+2)$-dimensional
Ricci-flat metrics on the tangent bundle of $S^{n+1}$ \cite{stenzel}.  It was
described in detail, in a notation close to that which 
we shall be using here, in
\cite{cvgilupo}.\footnote{The only change in notation is that we now take the 
index range for the $SO(n)$ subgroup of $SO(n+2)$ to be $1\le i\le n$
rather than $3\le i\le n+2$.}  Let $L_{AB}$, which are antisymmetric in the 
fundamental $SO(n+2)$ indices $A, B,\ldots$, 
be left-invariant 1-forms on the 
group manifold $SO(n+2)$, obeying the exterior algebra
\be
dL_{AB} = L_{AC}\wedge L_{CB}\,.
\ee
Splitting the indices $A=(i,n+1,n+2)$, the $L_{ij}$ are the left-invariant 
1-forms of the $SO(n)$ subgroup.  We make the definitions of the 1-forms
\be
\sigma_i \equiv L_{i,n+1}\,,\qquad \tilde\sigma_i \equiv L_{i,n+2}\,,\qquad
\nu\equiv L_{n+1,n+2}\,,\label{sigsnu}
\ee
which lie in the coset $SO(n+2)/SO(n)$.  They obey the algebra
\bea
d\sigma_i &=& \nu\wedge \tilde\sigma_i + L_{ij}\wedge \sigma_j\,,\qquad
d\tilde\sigma_i = -\nu\wedge\sigma_i + L_{ij}\wedge \tilde\sigma_j\,,\qquad
d\nu =- \sigma_i\wedge\tilde \sigma_i\,,\nn\\
dL_{ij} &=& L_{ik}\wedge L_{kj} -\sigma_i\wedge\sigma_j -
\tilde\sigma_i\wedge\tilde\sigma_j\,.
\eea
We then consider the metric
\be
ds^2 = d\xi^2 + a^2\, \sigma_i^2 + b^2\, \tilde\sigma_i^2 + c^2\, \nu^2\,,
\label{stenzelmet}
\ee
where $a$, $b$ and $c$ are functions of the radial coordinate $\xi$.
We define also the vielbeins
\be
e^0=d\xi\,,\qquad e^i=a\,\sigma_i\,,\qquad
e^{\tilde i}= b\, \tilde\sigma_i\,,\qquad e^{\tilde 0}= c\, \nu\,.
\label{vielbein}
\ee

  The spin connection, curvature 2-forms and the Ricci tensor are 
given in \cite{cvgilupo}.  It is also shown there that if one defines
a new radial coordinate $\eta$ such that $a^n\, b^n\, c\, d\eta=d\xi$,
then the Ricci-flat equations can be derived from the Lagrangian
$L=T-V$ where
\bea
T&=& \alpha'\, \gamma' + \beta'\, \gamma' + n\, \alpha'\, \beta'
  + \ft12(n-1)({\alpha'}^2 + {\beta'}^2)\,,\nn\\
V &=& \ft14 (ab)^{2n-2}\, (a^4+b^4+c^4-2a^2\, b^2 - 2n (a^2+b^2)c^2)\,,
\eea
and $a=e^\alpha$, $b=e^\beta$, $c=e^\gamma$.  

   Writing the Lagrangian as
$L=\ft12 g_{ij}\, (d\alpha^i/d\eta)\, (d\alpha^j/d\eta) -V$,
where $\alpha^i=(\alpha,\beta,\gamma)$, the potential $V$ can be written in 
terms of a superpotential $W$, as \cite{cvgilupo}
\be
V= -\ft12\, g^{ij}\, \fft{\del W}{\del\alpha^i}\, \fft{\del W}{\del\alpha^j}\,,
\qquad W= \ft12 (ab)^{n-1}\, (a^2+b^2+c^2)\,.\label{ricciflatsuperpot}
\ee
(For a systematic discussion of when superpotentials can be introduced
for the cohomogeneity-one Einstein equations, see \cite{danwan1,danwan2}.)
This implies that the Ricci-flat conditions are satisfied if the first-order
equations 
\be
\fft{d\alpha^i}{d\eta}= g^{ij}\, \fft{\del W}{\del\alpha^j}
\ee
are obeyed.  This leads to the first-order equations \cite{cvgilupo}
\be
\dot a = \fft1{2 bc}\,(b^2+c^2-a^2)\,,\qquad
\dot b= \fft1{2ac}\, (a^2 + c^2 - b^2)\,,\qquad
 \dot c = \fft{n}{2ab}\, (a^2 + b^2 -c^2)\,,\label{ricciflatfo}
\ee
where $\dot a$ means $da/d\xi$, etc.  

   These first-order equations are in fact the 
conditions that follow from requiring that the metrics be 
Ricci-flat and K\"ahler, namely that $R_{ab}=0$ and that
the K\"ahler form 
\be
J = -e^0\wedge e^{\tilde 0} + e^i\wedge e^{\tilde i} =
  -c\, d\xi\wedge \nu + a b\, \sigma_i\wedge \tilde\sigma_i\label{Jdef}
\ee
be covariantly constant.  In fact, they can be derived more simply by 
requiring 
\be
dJ=0\,,\qquad d\Omega_{n+1}=0\,,\label{ricciflatK}
\ee
where 
\be
\Omega_{n+1}\equiv \epsilon^0\wedge \epsilon^1\wedge\cdots\wedge \epsilon^{n}
\label{holoform}
\ee
is the holomorphic $(n+1)$-form and we have defined \cite{cvgilupo}
\be
\epsilon^0 \equiv -e^0 + \im e^{\tilde 0} = -d\xi + \im c \, \nu\,,\qquad
\epsilon^i \equiv 
  e^i + \im e^{\tilde i} = a\, \sigma_i + \im b\, \tilde\sigma_i
\,.
\ee

   It is easy to incorporate a cosmological constant $\Lambda$, 
so that the equations of motion become $R_{ab} = \Lambda\, g_{ab}$.  As
was shown in \cite{Dancer:2002cq} , this Einstein condition is satisfied if the 
first-order equations (\ref{ricciflatfo}) are modified to
\be
\dot a = \fft1{2 bc}\,(b^2+c^2-a^2)\,,\qquad
\dot b= \fft1{2ac}\, (a^2 + c^2 - b^2)\,,\qquad
 \dot c = \fft{n}{2ab}\, (a^2 + b^2 -c^2)-\Lambda\, ab\,.\label{einsteinfo}
\ee
These Einstein-K\"ahler 
first-order equations can also be derived by modifying the
Ricci-flat K\"ahler conditions (\ref{ricciflatK}) to
\be
dJ=0\,,\qquad D\Omega_{n+1}=0\,,
\ee
where $D$ is the $U(1)$ gauge-covariant exterior derivative
\be
D\equiv d - \im\, \Lambda\, A\,,
\ee
and $A$ is the K\"ahler 1-form potential, $J=dA$.  From (\ref{Jdef}) and
the equation $(ab)'=c$ that follows from $dJ=0$, it is easy to see that 
we can take
\be
  A= -a b\, \nu\,.
\ee

   The potential $V$ and 
superpotential $W$ appearing in
(\ref{ricciflatsuperpot}) should be modified in the $\Lambda\ne 0$ case to
\bea
V&=&  \ft14 (ab)^{2n-2}\, (a^4+b^4+c^4-2a^2\, b^2 - 2n (a^2+b^2)c^2
   + 4 \Lambda\,  a^2 b^2 c^2)\,,\nn\\
W&=& \ft12 (ab)^{n-1}\, (a^2+b^2+c^2) -\fft{\Lambda}{n+1}\, (ab)^{n+1}\,.
\eea
(The new superpotential for the special case $n=2$ was given in 
\cite{kuperstein}.) 

\section{$\CP^{n+1}$ Metrics in Stenzel Form}

   We may now consider solutions of the first-order system of equations
(\ref{einsteinfo}) for Einstein metrics of the Stenzel form.  It is easy to 
see that for each value of $n$ there is a solution of (\ref{einsteinfo})
given by
\be
a=\sin\xi\,,\qquad b=\cos\xi\,,\qquad c=\cos2\xi\,,\label{CPsol}
\ee
with cosmological constant $\Lambda=2(n+2)$.  (Of course, one can trivially
apply scalings to obtain other values of the cosmological constant.)  

  As we shall now show, the metric (\ref{stenzelmet}) with $a$, $b$ and
$c$ given by (\ref{CPsol}) is in fact the Fubini-Study metric on
$\CP^{n+1}$, written in a non-standard way.  To see this, we shall present
the generalisation of a construction of $\CP^2$ given in \cite{bougib},
extended now to an arbitrary even dimension $D=2n+2$.  

  Let $e_{n+1}$ and $e_{n+2}$ be an orthonormal pair of column vectors 
in $\R^{n+2}$, where 
\be
e_{n+1}=(0,0,\ldots,0,1,0)^T\,,\qquad
e_{n+2}=(0,0,\ldots,0,0,1)^T\,,\label{e1e2}
\ee
and let $R$ be an arbitrary element of $SO(n+2)$, which acts on $\R^{n+2}$
through matrix multiplication.  We then define the complex $(n+2)$-vector
\be
Z = R\, (\sin\xi\, e_{n+1} + \im\, \cos\xi\, e_{n+2})\,,\label{Zdef}
\ee
which clearly satisfies $Z^\dagger\, Z=1$.\footnote{Since $Z$ and $-Z$
are the same point in $\CP^{n+1}$, this means that when $n$ is
even (and hence $-R$ is in $SO(n+2)$ if $R$ is in $SO(n+2)$), the
group that acts effectively on $\CP^{n+1}$ is the 
projective special orthogonal group $PSO(n+2)=SO(n+2)/\Z_2$.  By
contrast, when $n$ is odd $SO(n+2)$ is centreless, and so the entire
$SO(n+2)$ acts effectively on $\CP^{n+1}$.}  The standard construction of
the Fubini-Study metric on $\CP^{n+1}$ is given, for $Z\in \C^{n+2}$ and 
satisfying $Z^\dagger\, Z=1$, by
\be
ds^2 = dZ^\dagger\, dZ - |Z^\dagger\, dZ|^2\,.
\label{fubistud}
\ee

   Defining the 1-forms $L_{AB}$ on $SO(n+2)$ by
\be
dR\, R^{-1} = \ft12 L_{AB}\, \widetilde M_{AB}\,,
\ee
where $\widetilde M_{AB}$ are the generators of the Lie algebra of $SO(n+2)$,
and introducing also the $SO(n+2)$-conjugated generators
\be
M_{AB}= R^T\, \widetilde M_{AB}\, R\,,
\ee
we see from (\ref{Zdef}) that
\be
dZ = R\, \Big[(L\cdot M)\, (\sin\xi\, e_{n+1} + \im \cos\xi\, e_{n+2}) +
               (\cos\xi\, e_{n+1} - \im\sin\xi\, e_{n+2})\, d\xi\Big]\,,
\ee
where we have defined $(L\cdot M)=\ft12 L_{AB}\, M_{AB}$.
We may take the generators $M_{AB}$ to have components given simply by
\be
(M_{AB})_{CD} = \delta_{AC}\, \delta_{BD}- \delta_{AD}\, \delta_{BC}\,,
\ee
and so we can choose a basis where 
$e_A^T\,  (L\cdot M)\, e_B = L_{AB}$.  Note that the $L_{AB}$
are left-invariant 1-forms of $SO(n+2)$.  It then follows
that
\be
Z^\dagger\, dZ = \im \sin2\xi\, L_{n+1,n+2}\,,\qquad
dZ^\dagger\, dZ = d\xi^2 -\sin^2\xi\, [(L\cdot M)^2]_{n+1,n+1} -
                  \cos^2\xi\, [(L\cdot M)^2]_{n+2,n+2}\,,
\ee
with
\bea
{[}(L\cdot M)^2{]}_{n+1,n+1} &=& L_{n+1,A} L_{A,n+1}= -(L_{n+1,n+2})^2 
    - (L_{i,n+1})^2\,,\nn\\
{[}(L\cdot M)^2{]}_{n+2,n+2} &=& L_{n+2,A} L_{A,n+ 2}= -(L_{n+1,n+2})^2 
   - (L_{i,n+2})^2\,.
\eea
In view of the definitions (\ref{sigsnu}), we therefore find that the
Fubini-Study metric (\ref{fubistud}) on $\CP^{n+1}$ can be written as
\be
ds^2 = d\xi^2 + \sin^2\xi\, \sigma_i^2 + \cos^2\xi\, \tilde\sigma_i^2 + 
   \cos^2 2\xi\, \nu^2\,,\label{CPstenmet}
\ee
which is precisely the metric we obtained above in (\ref{CPsol}).  

  The curvature 2-forms, which can be calculated from equations given in
\cite{cvgilupo}, turn out to be
\bea
\Theta_{0i}&=& e^0\wedge e^i - e^{\tilde 0}\wedge e^{\tilde i}\,,\qquad
  \Theta_{0\tilde i}= e^0\wedge e^{\tilde i} +
      e^{\tilde 0}\wedge e^i\,,\nn\\
\Theta_{0\tilde 0}&=& 4 e^0\wedge e^{\tilde 0} - 2e^i\wedge e^{\tilde i}\,,
\qquad \Theta_{ij}= e^i\wedge e^j + e^{\tilde i}\wedge e^{\tilde j}\,,\nn\\
\Theta_{\tilde i\tilde j} &=& e^{\tilde i}\wedge e^{\tilde j} + e^i\wedge e^j
\,,\qquad
\Theta_{i\tilde j} = e^i\wedge e^{\tilde j} -e^{\tilde i}\wedge e^j +
  2 (e^k\wedge e^{\tilde k} - e^0 \wedge e^{\tilde 0})\,
    \delta_{ij}\,,\nn\\
\Theta_{\tilde 0 i} &=&e^{\tilde 0}\wedge e^i + e^0\wedge e^{\tilde i} 
\,,\qquad
 \Theta_{\tilde 0\tilde i} =e^{\tilde 0}\wedge e^{\tilde i} - e^0\wedge e^i
\,,
\eea
where we are using the vielbein basis defined in (\ref{vielbein}).
The $\CP^{n+1}$ metrics are Einstein, with $R_{ab}= 2(n+2)\, g_{ab}$.
Note that as expected for the Fubini-Study metrics, the curvature has
constant holomorphic sectional curvature, and can be written as
\be
\Theta_{AB} = e^A\wedge e^B + J_{AC}\, J_{BD}\, e^C\wedge e^D +
    2 J_{AB}\, J\,,
\ee
where $J$ is the K\"ahler form, given in (\ref{Jdef}).

   It will also be useful for future reference to note that the 
$\CP^{n+1}$ metric (\ref{CPstenmet}) can be rewritten in terms of a 
new radial coordinate $\tau= \log\tan(\xi+\ft14\pi)$ as
\be
ds^2= \ft14\sech^2\tau\, d\tau^2 + \sinh^2\ft12\tau\, \sech\tau\, \sigma_i^2
+\cosh^2\ft12\tau\, \sech\tau\, \tilde\sigma_i^2 + \sech^2\tau\, \nu^2\,.
\label{CPstenmet2}
\ee
The radial coordinate ranges from $\tau=0$ at the $S^{n+1}$ bolt to
$\tau=\infty$ at the $SO(n+2)/(SO(n)\times SO(2))/\Z_2$ bolt.

\subsection{Global structure of the $\CP^{n+1}$ metrics}

The radial coordinate $\xi$ lies in the interval $0\le\xi\le\ft14\pi$.  
As $\xi$ goes to zero, the metric (\ref{CPstenmet}) 
extends smoothly onto a space that
has the local form $\R^{n+1}\times S^{n+1}$.  As can be seen by comparing with
the Ricci-flat Stenzel metrics as given in \cite{cvgilupo}, the metrics
take the same form in the vicinity of the origin.  The principal orbits 
when $0<\xi<\ft14 \pi$ are the Stiefel manifold $SO(n+2)/SO(n)$ divided by
$\Z_2$.  

At the other end of
the range of the $\xi$ coordinate, we see that as $\xi$ approaches 
$\ft14 \pi$, the metric (\ref{CPstenmet}) extends smoothly onto
$\R^2\times G_2(\R^{n+2})/\Z_2$.  The reason for the
$\Z_2$ quotient was discussed in footnote 3.  It is reflected in the  
fact that the integral $\oint\nu$ around the degenerate orbit at
$\xi=\pi/4$ must equal $\pi$, rather than $2\pi$.  This can be compared
with the situation in metrics discussed in section 5.1 below, for which
one has $\oint\nu=2\pi$ at the analogous degenerate orbit.

In the language of nuts and bolts, the $\CP^{n+1}$ manifold is 
described here in a form where there is an $S^{n+1}$ degenerate orbit 
or bolt at $\xi=0$ and
an $SO(n+2)/(SO(n)\times SO(2))/\Z_2$ bolt at $\xi=\ft14\pi$.

Since $\oint\nu=\pi$ at the bolt,
this implies that the level surfaces at fixed $\xi$ between the 
endpoints are the Stiefel manifold $SO(n+2)/SO(n)$ divided by $\Z_2$.

  Although the local form of the $\CP^{n+1}$ metrics near to the $S^{n+1}$
bolt at $\xi=0$ is similar to that of the Stenzel metrics on $T^* S^{n+1}$
near their $S^{n+1}$ bolt, the $\Z_2$ factoring of the 
$SO(n+2)/SO(n)$ principal orbits in the $\CP^{n+1}$ metrics that we discussed
above means that one cannot, strictly speaking, view the $\CP^{n+1}$
metrics as ``compactifications'' of the Stenzel metrics.  Rather, $\CP^{n+1}$
can be viewed as a ``compactification'' of the $\Z_2$ quotient  of the
Stenzel manifold.  As can be seen from the construction of the Stenzel
metrics given in section 2.1 of \cite{cvgilupo}, where the Stenzel manifold
is described by the complex quadric $z^a\, z^a=a^2$ in $\C^{n+2}$, one
can divide by $\Z_2$, with the action $\Z_2:\ z^a\rightarrow -z^a$, and
since this acts freely the quotient is still a smooth manifold.  As a
further cautionary remark, it should be noted that the Ricci-flat 
Stenzel metric on $T^* S^{n+1}$ does
not arise as a limit of the $\CP^{n+1}$ metric (\ref{CPstenmet}) in which
the cosmological constant is sent to zero.

\section{Other Exact Solutions of Stenzel Form}

  There are two other simple examples of Einstein metrics, with  a positive  cosmological constant
$\Lambda$, that take the Stenzel form (\ref{stenzelmet}), on the manifolds
$G_2(R^{n+3})$ and $S^{n+1}\times S^{n+1}$.  We present these  
in sections 5.1 and 5.2.  In section 5.3, by making 
appropriate analytic continuations,  we obtain Einstein metrics with
negative cosmological constant on non-compact forms of 
$\CP^{n+1}$, $G_2(\R^{n+3})$ and $S^{n+1}\times S^{n+1}$.

\subsection{Metrics on the Grassmannians $G_2(\R^{n+3})$}

  It is easy to see that the functions $a=\sin\xi$, 
$b=1$, $c=\cos\xi$ give a solution of 
the first-order equations (\ref{einsteinfo}), with $\Lambda= n+1$.
This gives another Einstein-K\"ahler metric,
\be
ds^2 = d\xi^2 + \sin^2\xi\, \sigma_i^2 + \tilde\sigma_i^2 
 + \cos^2\xi\, \nu^2\,.\label{mmetric}
\ee
The coordinate $\xi$ ranges from $0$ to $\ft12\pi$.  Near $\xi=0$ the
metric again looks locally like the Stenzel metric near its origin, and there 
is an $S^{n+1}$ bolt at $\xi=0$.  
The metric extends smoothly onto $\xi=\ft12\pi$,  provided that 
the integral $\oint\nu$ around $\xi=\pi/2$ is equal to $2\pi$.  Thus
in contrast to the $\CP^{n+1}$ metrics discussed in the previous section,
where we found that regularity at the degenerate orbit required $\oint\nu=\pi$
and hence implied the non-degenerate 
level surfaces were $SO(n+2)/SO(n)/\Z_2$, in the
present case the level surfaces are $SO(n+2)/SO(n)$.   The bolt at 
$\xi=\ft12\pi$ is the Grassmann manifold 
$G_2(\R^{n+2})=SO(n+2)/(SO(n)\times SO(2))$.

  For future reference, we note that here if we introduce a new radial 
coordinate defined by $\tau= 2\log\tan(\ft12\xi +\ft14\pi)$, the metric
(\ref{mmetric}) becomes
\be
ds^2= \ft14\sech^2\ft12\tau\, d\tau^2 + \tanh^2\ft12\tau\, \sigma_i^2 +
\tilde\sigma_i^2 + \sech^2\ft12\tau\, \nu^2\,.\label{mmetric2}
\ee
The radial coordinate ranges from the $S^{n+1}$ bolt at $\tau=0$
to the $G_2(\R^{n+2})$ bolt at $\tau=\infty$.

  The curvature 2-forms, which can again be calculated from equations given in 
\cite{cvgilupo}, turn out to be
\bea
\Theta_{0i}&=& e^0\wedge e^i + e^{\tilde 0}\wedge e^{\tilde i}\,,\qquad
  \Theta_{0\tilde i}=0\,,\nn\\
\Theta_{0\tilde 0}&=& e^0\wedge e^{\tilde 0} +e^i\wedge e^{\tilde i}\,,
\qquad \Theta_{ij}= e^i\wedge e^j + e^{\tilde i}\wedge e^{\tilde j}\,,\nn\\
\Theta_{\tilde i\tilde j} &=& e^{\tilde i}\wedge e^{\tilde j} + e^i\wedge e^j
\,,\qquad 
\Theta_{i\tilde j} = (e^k\wedge e^{\tilde k} + e^0 \wedge e^{\tilde 0})\,
    \delta_{ij}\,,\nn\\
\Theta_{\tilde 0 i} &=&0\,,\qquad 
 \Theta_{\tilde 0\tilde i} =e^{\tilde 0}\wedge e^{\tilde i} + e^0\wedge e^i
\,.\label{grasscurv}
\eea
(We have chosen the vielbein basis $e^0=d\xi$, $e^i=\sin\xi\, \sigma_i$, 
$e^{\tilde i}= \tilde\sigma_i$ and $e^{\tilde 0}=-\cos\xi\, \nu$ here.)  Note
that if we now define the indices
\be
I= (0,i)\,,\qquad \tilde I= (\tilde 0, \tilde i)\,,\qquad 0\le I\le n\,,
\ee
where $\tilde I=I+n+1$,
then the curvature 2-forms in (\ref{grasscurv}) can be written in the 
more compact form 
\be
\Theta_{IJ}= e^I\wedge e^J + e^{\tilde I}\wedge e^{\tilde J}\,,\qquad
\Theta_{\tilde I\tilde J}= e^{\tilde I}\wedge e^{\tilde J} 
           + e^{I}\wedge e^{J}\,,
\qquad
\Theta_{I\tilde J}= e^K\wedge e^{\tilde K}\, \delta_{IJ}\,.\label{grasscurv2}
\ee
 From this it can be seen that the metrics (\ref{mmetric}) are Einstein,
with $R_{ab}= (n+1)\, g_{ab}$.

   The metrics (\ref{mmetric}) are in fact metrics on
the Grassmannian manifolds
\be
G_2(\R^{n+3})=\fft{SO(n+3)}{SO(n+1)\times SO(2)}\,.\label{Gcoset}
\ee
This can be seen by starting from the
left-invariant 1-forms $\hat L_{AB}$ of $SO(n+3)$, with $0\le A\le n+2$,
decomposing the indices as
$A=(I,a)$, where $I=0,\ldots,n$ and $a=n+1,n+2$, and then defining 
the $2(n+1)$-bein
\be
e^I=\hat L_{I,n+1}\,,\qquad e^{\tilde I}= \hat L_{I,n+2}\,,
\ee
for the metric $ds^2=e^I\otimes e^I + e^{\tilde I}\otimes e^{\tilde I}$,
where $\tilde I = I+n+1$.
The spin connection is then given by
\be
\omega_{IJ}= -\hat L_{IJ}\,, \qquad \omega_{\tilde I\tilde J}= -\hat L_{ij}\,,
\qquad \omega_{I\tilde J}= -\delta_{IJ}\, \hat L_{n+1,n+2}\,,
\ee
and hence the curvature 2-forms are
\be
\Theta_{IJ}=e^I\wedge e^J +e^{\tilde I}\wedge e^{\tilde J}\,,\qquad
\Theta_{\tilde I\tilde J}= e^{\tilde I}\wedge e^{\tilde J} +
  e^I\wedge e^J\,,\qquad
\Theta_{I\tilde J}= e^K\wedge e^{\tilde K}\, \delta_{IJ}\,.
\ee
Thus the curvature for these metrics on the Grassmannian manifolds
$G_2(\R^{n+3})$ is in precise agreement with the curvature (\ref{grasscurv2})
that we found for the metrics (\ref{mmetric}). 
It is easily verified that the 2-form
\be
J= e^I\wedge e^{\tilde I}
\ee
is closed, and furthermore covariantly constant, and hence it is
a K\"ahler form for $G_2(\R^{n+3})$.

   Since the metrics (\ref{mmetric}) are locally similar to the 
Ricci-flat Stenzel metrics near the $S^{n+1}$ bolt at $\xi=0$, and
the principal orbits for $\xi>0$ are the Stiefel manifolds
$SO(n+2)/SO(n)$, just as in the Ricci-flat Stenzel metrics, one may
view the metrics (\ref{mmetric}) as a kind of ``compactification'' of the
Stenzel metrics. However, as we remarked earlier in the
context of the $\CP^{n+1}$ metrics, one should view this interpretation with
some caution, since there is no $\Lambda\rightarrow 0$ limit of the
the metrics (\ref{mmetric}) that gives the Ricci-flat Stenzel metrics.

\subsection{An $S^{n+1}\times S^{n+1}$
 solution of the second-order equations}\label{sxssection}

   We can also find a solution of the second-order Einstein equations
that is not a solution of the first-order equations (\ref{einsteinfo}),
and thus it is not K\"ahler (at least with respect to the almost complex
structure defined by $J$ in (\ref{Jdef})).
This is given by
\be
a= \sin\xi\,,\qquad b=\cos\xi\,,\qquad c=1\,,\label{sxsmet}
\ee
and it is Einstein with $\Lambda= 2n$.  This is in fact the
standard product metric on $S^{n+1}\times S^{n+1}$.  This can be
seen by introducing two orthonormal vectors in $\R^{n+2}$, as in (\ref{e1e2}),
and then defining the two real $(n+2)$-vectors
\be
X = R\, (\sin\xi\, e_{n+1} + \cos\xi\, e_{n+2})\,,\qquad
Y= R\, (\sin\xi\, e_{n+1} - \cos\xi\, e_{n+2})\,,
\ee
where $R$ is again a general element of $SO(n+2)$.  Note that these
satisfy $X^T \,X=1$ and $Y^T\, Y=1$.
The suitably scaled metric on $S^{n+1}\times S^{n+1}$ can be written as
\be
ds^2 = \ft12 dX^T\, dX + \ft12 dY^T\, dY\,.
\ee
Following analogous steps to those we used in the $\CP^{n+1}$ case, 
we find that here the metric on $S^{n+1}\times S^{n+1}$ becomes
\be
ds^2 = d\xi^2 + \sin^2\xi\, \sigma_i^2 + 
  \cos^2\xi\, \tilde\sigma_i^2 + \nu^2\,,\label{sxsmet2}
\ee
which is precisely the one given by (\ref{sxsmet}).  

 The coordinate $\xi$ here ranges over $0\le \xi\le \ft12\pi$.  The
metric has Stenzel-like behaviour near each endpoint, and can be viewed
as a flow from an $S^{n+1}$ bolt at one end to a ``slumped'' $S^{n+1}$ 
bolt at the other end.

  The curvature 2-forms are given by
\bea
\Theta_{0i}&=& e^0\wedge e^i + e^{\tilde 0}\wedge e^{\tilde i}\,,\qquad
  \Theta_{0\tilde i}= e^0\wedge e^{\tilde i} + e^{\tilde 0}\wedge e^i\,,\nn\\
\Theta_{0\tilde 0}&=& 0\,,
\qquad \Theta_{ij}= e^i\wedge e^j + e^{\tilde i}\wedge e^{\tilde j}\,,\nn\\
\Theta_{\tilde i\tilde j} &=& e^{\tilde i}\wedge e^{\tilde j} + e^i\wedge e^j
\,,\qquad
\Theta_{i\tilde j} = e^i\wedge e^{\tilde j} + e^{\tilde i}\wedge e^j\,,\nn\\
\Theta_{\tilde 0 i} &=& e^{\tilde 0}\wedge e^i + e^0\wedge e^{\tilde i}
\,,\qquad
 \Theta_{\tilde 0\tilde i} =e^{\tilde 0}\wedge e^{\tilde i} +e^0\wedge e^i
\,,\label{sxscurv}
\eea
from which it can be seen that the metrics are Einstein, 
with $R_{ab}=2n\, g_{ab}$.

\subsection{Non-compact manifolds with negative-$\Lambda$ Einstein metrics}

   By performing straightforward analytic continuations we can obtain
Einstein metrics with negative cosmological constant on non-compact 
forms of all three classes of manifolds that we have considered in this
paper.  The procedure is the same in all three cases, and comprises the
following steps.  First, we perform a Wick rotation of the cohomogeneity-one
coordinate $\xi$, sending $\xi\rightarrow \im\, \xi$. Next, we perform a
Wick rotation on the coordinates $x^A$ of the $\R^{n+2}$ Euclidean space, 
sending $x^{n+2}\rightarrow \im\, x^{n+2}$.  This has the effect of sending
\be
\sigma_i\longrightarrow \sigma_i\,,\qquad \tilde\sigma_i\longrightarrow
\im\, \tilde\sigma_i\,,\qquad \nu \longrightarrow \im\, \nu\,.
\ee
Finally, we reverse the sign of the metric.  The metrics (\ref{CPstenmet}),
(\ref{mmetric}) and (\ref{sxsmet2}) then become
\bea
\widetilde{\CP^{n+1}}:&&\qquad ds^2= d\xi^2 + \sinh^2\xi\, \sigma_i^2 + 
  \cosh^2 \xi\, \tilde\sigma_i^2 + \cosh^2 2\xi\, \nu^2\,,\nn\\
\widetilde{G_2(R^{n+3})}:&& \qquad ds^2 =d\xi^2 + \sinh^2\xi\, \sigma_i^2 + 
\tilde\sigma_i^2 + \cosh^2\xi\, \nu^2\,,\nn\\
H^{n+1}\times H^{n+1}:&&\qquad ds^2 = d\xi^2 + \sinh^2\xi\, \sigma^2_i +
  \cosh^2\xi\, \tilde\sigma_i^2 + \nu^2\,,\label{noncompmet}
\eea
where $\widetilde{\CP^{n+1}}$ and $\widetilde{G_2(R^{n+3})}$ denote the
non-compact forms of $\CP^{n+1}$ and $G_2(\R^{n+3})$, and $H^{n+1}$ denotes
the hyperbolic space that is the non-compact form of $S^{n+1}$.
The left-invariant 1-forms $\sigma_i$, $\tilde\sigma_i$ and $\nu$,
which now span the coset $SO(n+1,1)/(SO(n)\times SO(1,1))$, 
satisfy the exterior algebra
\bea
d\sigma_i &=& -\nu\wedge \tilde\sigma_i + L_{ij}\wedge \sigma_j\,,\qquad
d\tilde\sigma_i = -\nu\wedge\sigma_i + L_{ij}\wedge \tilde\sigma_j\,,\qquad
d\nu =- \sigma_i\wedge\tilde \sigma_i\,,\nn\\
dL_{ij} &=& L_{ik}\wedge L_{kj} -\sigma_i\wedge\sigma_j +
\tilde\sigma_i\wedge\tilde\sigma_j\,.
\eea
The cosmological constants for the three metrics in (\ref{noncompmet})
are given by $\Lambda=-2(n+2)$, $\Lambda= -(n+1)$ and $\Lambda=-2n$,
respectively.  In each case the coordinate $\xi$ ranges from $\xi=0$
at the $S^{n+1}$ bolt to $\xi=\infty$.

\section{Six Dimensions} 

  The case of six dimensions, corresponding to $n=2$, is of particular
interest for a variety of applications in string theory.  In this case the
numerator group in the coset $SO(n+2)/(SO(n)\times SO(2))$ of the level 
surfaces of the Stenzel construction is $SO(4)$, which is (locally) the
product $SU(2)\times SU(2)$.  In this section we introduce Euler angles and
discuss their coordinate ranges.  We also make a comparison of our
six-dimensional exact solutions with the numerical results obtained in
\cite{kuperstein}.

\subsection{Euler angles and fundamental domains}

The left-invariant $SO(4)$ 1-forms $L_{AB}$
are related to two sets of left-invariant $SU(2)$ 1-forms 
$\Sigma_i$ and $\widetilde\Sigma_i$ according to
\bea
\Sigma_1&=& L_{23} + L_{14}\,,\qquad \Sigma_2=L_{31} + L_{24}\,,\qquad
\Sigma_3= L_{12} + L_{34}\,,\nn\\
\widetilde\Sigma_1&=& L_{23} - L_{14}\,,\qquad 
\widetilde\Sigma_2=L_{31} - L_{24}\,,\qquad
\widetilde\Sigma_3= L_{12} - L_{34}\,.
\eea
These therefore satisfy
\be
d\Sigma_i = -\ft12 \epsilon_{ijk}\, \Sigma_j\wedge\Sigma_k\,,\qquad
d\widetilde\Sigma_i = -\ft12 \epsilon_{ijk}\, \widetilde\Sigma_j
       \wedge\widetilde\Sigma_k\,.
\ee
In view of the definitions (\ref{sigsnu}), we therefore have that 
\bea
\sigma_1&=& -\ft12(\Sigma_2+\widetilde\Sigma_2)\,,\qquad
\sigma_2= \ft12 (\Sigma_1 +\widetilde\Sigma_1)\,,\nn\\
\tilde\sigma_1&=& \ft12(\Sigma_1 - \widetilde\Sigma_1)\,,\qquad
\tilde\sigma_2= \ft12 (\Sigma_2 -\widetilde\Sigma_2)\,,\nn\\
\nu &=& \ft12(\Sigma_3 -\widetilde\Sigma_3)\,,
  \qquad L_{12} =\ft12(\Sigma_3 + \widetilde\Sigma_3)\,.
\eea

    The $SU(2)$ left-invariant 1-forms $\Sigma_i$ and $\widetilde\Sigma_i$
may be parameterised in terms of Euler angles $(\theta,\phi,\psi)$ and
$(\tilde\theta,\tilde\phi,\widetilde\psi)$ in the standard way:
\crampest
\bea
\Sigma_1 &\!\!\!\! =&\!\!\!\! 
\sin\psi\,\sin\theta\, d\phi\ +\cos\psi\, d\theta\,,\
\Sigma_2= \cos\psi\,\sin\theta\, d\phi -\sin\psi\, d\theta\,,\
\Sigma_3 = d\psi+\cos\theta\, d\phi\,,\nn\\
\widetilde\Sigma_1 &\!\!\!\!=&\!\!\!\!
\sin\widetilde\psi\,\sin\tilde\theta\, d\tilde\phi+\cos\widetilde\psi\, 
d\tilde\theta \,,\
\widetilde\Sigma_2= \cos\widetilde\psi\,\sin\tilde\theta\, d\tilde\phi
 -\sin\widetilde\psi\, d\tilde\theta \,,\
\widetilde\Sigma_3 = d\widetilde\psi+\cos\tilde\theta\, d\tilde\phi\,. 
\eea
\uncramp

   There are four inequivalent connected Lie groups whose Lie algebra is
$\mathfrak{so}(4)$, namely 
\be
SU(2)\times SU(2)\,,\qquad SO(4)\,,\qquad SU(2)\times SO(3)\,,
\qquad SO(3)\times SO(3)\,.
\ee
These are distinguished by their fundamental domains in the 
$(\psi,\widetilde\psi)$ plane.  We have
\bea
SU(2)\times SU(2):&& 0\le\psi<4\pi\,,\qquad 0\le \widetilde\psi <4\pi\,,\nn\\
SO(4):&& 0\le\psi<4\pi\,,\qquad 0\le \widetilde\psi <4\pi\,,\quad
\hbox{and}\quad (\psi,\widetilde\psi)\equiv (\psi+2\pi,\widetilde\psi+2\pi)\,,
\nn\\
SU(2)\times SO(3):&& 0\le\psi<4\pi\,,\qquad 0\le \widetilde\psi <2\pi\,,\nn\\
SO(3)\times SO(3):&& 0\le\psi<2\pi\,,\qquad 0\le \widetilde\psi <2\pi\,.
\eea
These identifications can be expressed in terms of the following 
generators:
\bea
T:&& (\psi,\widetilde\psi)\longrightarrow (\psi+4\pi,\widetilde\psi)\,,\nn\\
\widetilde T:&& (\psi,\widetilde\psi)\longrightarrow 
         (\psi,\widetilde\psi+4\pi)\,,\nn\\
  S:&& (\psi,\widetilde\psi)\longrightarrow
         (\psi+2\pi ,\widetilde\psi)\,,\nn\\
\widetilde S:&& (\psi,\widetilde\psi)\longrightarrow 
         (\psi,\widetilde\psi+2\pi)\,,\nn\\
D:&& (\psi,\widetilde\psi)\longrightarrow
         (\psi+2\pi,\widetilde\psi+2\pi)\,.
\eea
Clearly, these all commute, and they obey
\be
S^2=T\,,\qquad \widetilde S^2 =\widetilde T\,,\qquad D^2 = T\widetilde T\,.
\ee
Starting from $(\psi,\widetilde \psi)$ defined in $\R^2$, the four
groups are obtained by quotienting by the action of the generators listed 
below:
\bea
SU(2)\times SU(2):&& T\,,\qquad \widetilde T\,,\nn\\
SO(4):&& T\,, \qquad \widetilde T\,,\qquad D\,,\nn\\
SU(2)\times SO(3):&& T\,,\qquad \widetilde S\,,\nn\\
SO(3)\times SO(3):&& S\,,\qquad \widetilde S\,.
\eea

Defining the oblique coordinates
\be
\psi_\pm = \psi \pm \widetilde\psi\,,
\ee
the fundamental domains given above for the four cases can be re-expressed 
in terms of $\psi_+$ and $\psi_-$.  This can be done straightforwardly by
plotting the domain in the $(\psi,\widetilde\psi)$ plane, partitioning 
where necessary into triangular sub-domains, and acting with the
appropriate translation generators listed above in order to achieve a 
connected fundamental domain in the $(\psi_+,\psi_-)$ plane.  This gives
\bea
SU(2)\times SU(2):&& 0\le\psi_+ <8\pi\,,\qquad 0\le \psi_- <4\pi\,,\nn\\
SO(4):&& 0\le\psi_+ <4\pi\,,\qquad 0\le \psi_- <4\pi\,,\nn\\
SU(2)\times SO(3):&& 0\le\psi_+ <8\pi\,,
\qquad 0\le \psi_- <2\pi\,,\nn\\
SO(3)\times SO(3):&& 0\le\psi_+<4\pi\,,\qquad 0\le\psi_- <2\pi\,.
\label{periods}
\eea

  Consider first the $\CP^3$ metric, given by (\ref{CPstenmet}) with
$n=2$.  Near the upper endpoint of the coordinate $\xi$, at $\xi=\pi/4$, we 
may define $\xi=\pi/4-\alpha$, and the metric
approaches
\be
ds^2 \rightarrow d\alpha^2 + \alpha^2\, (d\psi_- +\cos\theta\, d\phi-
   \cos\tilde\theta\, d\tilde\phi)^2 + \ft12 (\sigma_i^2 + \tilde\sigma_i^2)
\,.
\ee
This extends smoothly onto $\alpha=0$ provided that $\psi_-$ is assigned the
period
\be
\Delta\psi_- = 2\pi\,.\label{so4z2}
\ee
Comparing with the periodicity conditions in (\ref{periods}) for $SO(4)$, we
see that the $SO(4)$ group manifold is factored by $\Z_2$. This is consistent
with the fact that $Z$, defined by (\ref{Zdef}) is equivalent to $-Z$ in
$\CP^3$:  Since $-R$ is in $SO(4)$ if $R$ is in $SO(4)$, we should
identify $R$ and $-R$ in the construction (\ref{Zdef}), and hence we should
impose (\ref{so4z2}).  This identification divides $SO(4)$ by its $\Z_2$
centre, giving the projective special orthogonal group $PSO(4) =
   SO(3)\times SO(3)$.  Thus the principal orbits are $V_2(\R^4)/\Z_2$,
where $V_2(\R^4)$ is the Stiefel manifold $SO(4)/SO(2)$. 

   Turning now to the metric on the six-dimensional Grassmannian 
manifold $G_2(\R^5)=SO(5)/(SO(3)\times SO(2))$, given by 
(\ref{mmetric}) with $n=2$, we see that near the upper end of the
range of the $\xi$ coordinate, at $\xi=\pi/2$, the metric takes the
form
\be
ds^2 \rightarrow d\alpha^2 + \ft14 \alpha^2\,
(d\psi_- +\cos\theta\, d\phi-
   \cos\tilde\theta\, d\tilde\phi)^2 + \sigma_i^2 + \tilde\sigma_i^2\,,
\ee
where we have written $\xi=\pi/2 -\alpha$.  The metric
extends smoothly onto $\alpha=0$ provided that $\psi_-$ has the
period
\be
\Delta\psi_-=4\pi\,,
\ee
and so from (\ref{periods}) we see that in this case the group acting on the 
$\xi=\,$constant level surfaces is precisely $SO(4)$. The principal
orbits are the Stiefel manifold $V_2(\R^4)=SO(4)/SO(2)$, which is 
often called $T^{1,1}$.

  Finally, in the case of the $S^3\times S^3$ metric given by (\ref{sxsmet2}) 
with $n=2$, it is evident from the general construction described in 
section \ref{sxssection} that the group acting on the level surfaces 
should be precisely $SO(n+2)$, and thus when $n=2$ we should have 
$\Delta\psi_-=4\pi$.  This can by confirmed by noting from (\ref{sxscurv})
that the metrics (\ref{sxsmet2}) satisfy $R_{ab}= 2n\, g_{ab}$ and thus
when $n=2$ it must be isomorphic to the product metric on two 3-spheres 
of radius $1/\sqrt2$.  Calculating the volume using the metric (\ref{sxsmet2})
then confirms that indeed we must have $\Delta\psi_-=4\pi$.  The principal
orbits are the Stiefel manifold $V_2(\R^4)=SO(4)/SO(2)$.

\subsection{Comparison with numerical solution in \cite{kuperstein}}

   A solution of the first-order equations (\ref{einsteinfo}) in six
dimensions was obtained recently by Kuperstein \cite{kuperstein}.  The 
left-invariant 1-forms on the five-dimensional principal orbits were
denoted by $(g_1,g_2,g_3,g_4,g_5)$ in \cite{kuperstein}, and one can show
that these may be related to our 1-forms by
\be
g_1=\fft1{\sqrt2}\, \sigma_1\,,\quad
g_2=\fft1{\sqrt2}\, \sigma_2\,,\quad
g_3=-\fft1{\sqrt2}\, \tilde\sigma_2\,,\quad
g_4=\fft1{\sqrt2}\, \tilde\sigma_1\,,\quad
g_5=2\nu\,.
\ee
Comparing the metric given in eqn (2.1) of \cite{kuperstein} with our
metric (\ref{stenzelmet}), we see that the metric functions $e^w$, $e^y$
and $e^z$ in \cite{kuperstein} are related to our metric functions
$a$, $b$ and $c$ by
\be
e^w=\ft32 a^2 b^2 c^2\,,\qquad e^y= \fft{a}{b}\,,\qquad e^z= 2 a b\,.
\ee
The radial variable used in \cite{kuperstein} is the same as the $\tau$
variable that we introduced in the rewriting of the $\CP^{n+1}$ metrics
(\ref{CPstenmet2}) and the $G_2(\R^{n+3})$ metrics (\ref{mmetric2}).
Note that both for our $\CP^3$ and our $G_2(\R^5)$ metrics, we have
$e^y=\tanh\ft12\tau$, as in \cite{kuperstein}.

  It is now a simple matter to compare the asymptotic forms of the 
metric functions found in the numerical solution in \cite{kuperstein}
with those for the exact solutions we have obtained in this paper.  In
particular, we see that near $\tau=0$ the function $e^z$ takes the form
\bea
\CP^3:&& \qquad e^z = \tanh\tau = \tau -\ft13 \tau^3 + \cdots\,,\nn\\
G_2(\R^5):&&\qquad e^z = 2\tanh \ft12\tau= \tau-\ft1{12}\, \tau^2+\cdots\,.
\eea
Ar large $\tau$, we have 
\bea
\CP^3:&& \qquad e^z = \tanh\tau = 1- 2 e^{-2\tau} + 2 e^{-4\tau} 
   +\cdots\,,\nn\\
G_2^(\R^5):&&\qquad e^z=2\tanh\ft12\tau = 2( 1 - 2 e^{-\tau} + 
  2 e^{-2\tau}+\cdots )\,.
\eea
Comparing with the asymptotic forms given in eqns (3.4) and (3.5) of
\cite{kuperstein}, we see that the metric that was found numerically there 
coincides with our exact solution for the Einstein-K\"ahler 
metric on the Grassmannian
manifold $G_2(\R^5)=SO(5)/(SO(3)\times SO(2))$, with the scale size
$R=\sqrt{2/3}$, and the expansion coefficients 
$C_{IR}=1$, and $C_{UV}=-2$.  Of course, one can trivially
rescale our metric to obtain any desired value for $R$.\footnote{The 
coefficient $C_{UV}$ associated with the large-$\tau$ expansion 
in \cite{kuperstein}
is said to be approximately $+1.96$ in that paper, but clearly, given the  
form of the asymptotic expansion $e^z=3 R^2(1 + C_{UV}\, e^{-\tau} +
   \ft12 C_{UV}^2\, e^{-2\tau}+\cdots)$ appearing there, $C_{UV}$ must be
negative rather than positive, since $e^z$ approaches $3R^2$ from below
rather than above, as $\tau$ goes to infinity.}

\section{Conclusions}

In this paper we constructed three classes of exact  Einstein metrics of 
cohomogeneity one in $(2n+2)$ dimensions. These are  generalisations of 
the Stenzel construction of Ricci-flat metrics, in which a positive 
cosmological constant is introduced. We also studied the global structure 
of the  manifolds onto which these local metrics extend. 

\begin{itemize}

\item   The first class of metrics, which satisfy the first-order 
Stenzel equations  with a positive cosmological constant, are  therefore
Einstein-K\"ahler.  We demonstrated that these metrics are the standard 
Fubini-Study metrics on  the 
complex projective spaces $\CP^{n+1}$, though presented in an unusual form. 
The study of the global structure  revealed that  the principal orbits  
of these metrics are the Stiefel  manifolds $V_2(\R ^{n+2})=SO(n+2)/SO(n)$ 
of 2-frames in $\R^{n+1}$, quotiented by $\Z_2$.   
As the cohomogeneity-one coordinate approaches
$\xi=0$,  there is an $S^{n+1}$-dimensional  degenerate orbit or  bolt,
while at   $\xi=\ft14\pi$ there is an $SO(n+2)/(SO(n)\times SO(2))/\Z_2$ 
degenerate orbit.
The  special  case $n=1$, giving $\CP^2$, corresponds to a  solution  
obtained by a  geometrical construction in \cite{bougib}. 

\item   The second class  of metrics are also exact solutions of the 
first-order Stenzel 
equations with a positive cosmological constant.  These homogeneous 
Einstein-K\"ahler metrics extend
smoothly onto the Grassmannian manifolds 
$G_2(\R^{n+3})=SO(n+3)/((SO(n+1)\times
SO(2))$  of  oriented 2-planes in $\R^{n+3}$, whose principal orbits 
are again the Stiefel manifolds 
$V_2(\R^{n+2})$ (not factored by $\Z_2$ in this case),  
viewed as $U(1)$ bundles over the Grassmannian 
manifolds  $G_2(\R^{n+2})$. The metric has 
an $S^{n+1}$ bolt at $\xi=0$, and an
 $SO(n+2)/(SO(n)\times SO(2))$ bolt at $\xi=\ft12\pi$. 
The case $n=2$,
giving  $G_2(\R^5)=SO(5)/(SO(3)\times SO(2))$, 
is a solution that was found numerically
 in \cite{kuperstein}. It  represents a generalisation of the conifold 
metric (6-dimensional Stenzel metric) to include a positive  
cosmological constant.

 \item   The third class of  Einstein metrics does not, in 
general, satisfy the first-order equations.  A geometrical construction 
for these metrics demonstrates that they
extend smoothly onto the product manifolds $S^{n+1}\times S^{n+1}$. 
The  $n=1$ case was
first constructed  in \cite{Pope} and described  in detail
in appendix B of \cite{Cvetic:2002kj}.  This is  also the only case 
in this class where the Einstein metric is also K\"ahler. 

\end{itemize}

  By making appropriate analytic continuations, we also obtained
Einstein metrics with negative cosmological constant 
on non-compact forms of $\CP^{n+1}$, $G_2(\R^{n+3})$ 
and $S^{n+1}\times S^{n+1}$.

The compact Einstein spaces 
presented in this paper should play an important role in 
further studies of consistent compactifications of M-theory and 
string theory, as well in the context of the
AdS/CFT correspondence. In addition to the  $\CP^{3}$ metric in the 
Stenzel form, the role of the other classes of metrics with $n=2$, 
as well as those with $n>2$, deserves further investigation. 

K\"ahler, but not Ricci flat, metrics on the deformed conifold arise  in the theory of moduli space 
of $\CP^1$ lumps \cite{Speight:2001he,Speight:2001hq}. The method developed in this paper should be applicable to those metrics as well.

In this paper we also treated  the geometrical approach to 
quantum mechanics,  where  $\CP^{n}$ is the space of physically-distinct 
quantum states of a system with an $(n+1)$-dimensional Hilbert space, 
thus employing the  geometry of its Fubini-Study metric 
\cite{Kibble,Gibbons:1991sa,Brody:1999cw}.  The calculations
involving $\CP^2$  in  \cite{bougib}  were aimed at  evaluating 
the   Aharonov-Anandan phase for a 3-state
spin-1  system.  We have elaborated  
further on the formalism, and spelled out applications to the discussion of
quantum entanglement for systems with two qubits and three qubits.  
A  linear superposition of two unentangled states is in general entangled. 
The set of such bi-partite
states is spanned by the set of physically-distinct
unentangled product states, which
form the complex sub-variety    $\CP^1\times \CP^1 \subset\CP^3$,   
given in section 2.2 as an explicit Segre embedding. 

   The notion of entanglement depends on the factorisation of the total 
Hilbert space. In  the case of two qubits  the Hilbert space could be 
split into a product of two orthogonal two-dimensional subspaces, 
forming a complex Grassmannian manifold 
$G_2(\C^2)=SU(4)/(SU(2)_1\times SU(2)_2)$.  An example of that type is a 
two qubit system consisting of  
nucleons, with  $SU(2)_{1}$ and $SU(2)_2$ playing the role of 
isospin and spin symmetry
respectively.  

    Studies of quantum entanglement for more complex systems,
such as a  
three qubit system,
are  of great current interest, 
and the proposed 
geometric approach could shed further light on these important questions.
Within this context, we studied the tripartite quantum entanglement of
qubits, showing the vanishing of the Cayley hyperdeterminant.

Another area where the geometry of $\CP^n$ comes to the aid of physics
is in quantum control theory \cite{Brody:2014jaa}.

\vskip 0.5in
{\noindent\large  \bf Acknowledgments}
\vskip 0.1in
\noindent The work of M.C. 
is supported in part by the  DOE (HEP) Award DE-SC0013528, the 
Fay R. and Eugene L. Langberg Endowed Chair (M.C.) and the Slovenian 
Research Agency (ARRS).
The work of G.W.G. was supported in part by the award
of a LE STUDIUM Professorship held at the L.M.P.T. of the
University of Francois Rabelais.
The work of C.N.P.
is supported in part by DOE grant DE-FG02-13ER42020. 
M.C. thanks the Cambridge Centre for Theoretical Cosmology, and
G.W.G. and C.N.P. thank the  UPenn Center for Particle Cosmology, 
for hospitality  during the course of this work.

\end{document}